\documentclass[twocolumn]{revtex4}
\usepackage[utf8]{inputenc}
\usepackage{graphicx}

\begin{document}

\title{Electric susceptibility of antiferromagnetic multiferroics with cycloidal spin order at magnetoelectric effect associated with collinear component of spins}

\author{Pavel A. Andreev}
\email{andreevpa@physics.msu.ru}
\affiliation{Department of General Physics, Faculty of physics, Lomonosov Moscow State University, Moscow, Russian Federation, 119991.}


\begin{abstract}
The contribution of magnetoelectric effect to Landau--Lifshitz-Gilbert equation is considered in case when medium polarization is caused by parallel component of neighboring spins.
The result is presented for ferromagnetic and antiferromagnetic materials.
A comparison is given with the contribution of magnetoelectric effect to Landau--Lifshitz-Gilbert equation when medium polarization is caused by perpendicular component of spins. The dispersion dependence of spin waves in antiferromagnets with cycloidal equilibrium spin order is derived. The electric susceptibility and permittivity of antiferromagnetic multiferroics in which magnetoelectric effect is caused by collinear component of spins is obtained analytically.
\end{abstract}


\maketitle



\section{Introduction}

Collective excitations are fundamental phenomena in solid-state physics and other continuous media.
Most collective excitations or their corresponding quasi-particles are considered as perturbations of homogeneous equilibrium states.
In multiferroics, in media exhibiting mutual influence of magnetic, dielectric, elastic and other properties,
conditions arise for the formation of periodic equilibrium structures of magnetization. Spin waves in such systems are insufficiently explored
(see \cite{Fishman PRB 19}, where this issue is partially considered). Therefore, in this paper, the dispersion dependence of spin waves in antiferromagnetic multiferroics with the cycloidal spin order is analytically considered.

The study of the dielectric response of antiferromagnetic multiferroics of the second kind,
i.e. media in which the occurrence or change of the electric dipole moment is due to spin effects,
led to the experimental discovery of new collective excitations called electromagnons
\cite{Pimenov NP 06}. Therefore, along with the dispersion dependence of spin waves in periodic equilibrium structures,
it is important to obtain a connection between the dynamic polarization of the medium and perturbations of the electric field.
There are several mechanisms for the formation of polarization of the medium by spin effects
\cite{Tokura RPP 14}.
In this paper, one of the known mechanisms caused by the collinear component of spins is considered.
The corresponding electric dipole moment of a group of magnetic and non-magnetic ions is presented in the review
\cite{Tokura RPP 14} (see Fig. 2), and the macroscopic polarization is presented, for example, in the review
\cite{Dong AinP 15} (see equations 3, 9 and 10).
A spin-current model of the formation of the electric dipole moment of multiferroics for systems with non-collinear spins is known in the literature
(see
\cite{Katsura PRL 05} and review
\cite{Tokura RPP 14} Fig. 2).
Later, in the works
\cite{AndreevTrukh JETP 24} and \cite{AndreevTrukh PS 24}, a generalization of the spin-current model is proposed.
First, the proportionality coefficient is established between the polarization of the medium and the spin current.
Second, it is shown that the spin-current model leads to the formation of the electric dipole moment of multiferroics for systems with collinear spins.
It is deduced that the electric dipole moment of multiferroics with collinear spins is due to the Dzyaloshinsky-Moriya interaction.
A generalization of the formula for macroscopic polarization
\cite{AndreevTrukh PS 24} known from the discussion in the work
\cite{Dong AinP 15} is derived, namely, an additional term associated with the spatial inhomogeneity of the polarization was obtained.
This generalization is considered for both ferromagnets and antiferromagnets \cite{AndreevTrukh PS 24}.
As shown below, this additional term that leads to a non-zero moment of force in the Landau-Lifshitz equation is used in this work to calculate the permittivity.

Examples of calculating phase diagrams of antiferromagnets can be found in the work \cite{Gareeva PRB 13}.
There, the possibility of the existence of equilibrium structures of a more complex type, in comparison with those considered by us below, is shown.
However, their equilibrium properties are studied there, and perturbations are considered only for phases with the collinear order,
which nevertheless allows to detect instability conditions and estimate the boundaries of these phases.

Note that in the works \cite{Bychkov SSP 12} (dispersion dependence of spin and electromagnetic waves on the background of a ferromagnetic spiral structure),
\cite{Bychkov JMMM 13}
(acoustic waves are additionally taken into account, including the acoustic Faraday effect)
spin waves on the background of spiral magnetic equilibrium structures are already considered.
However, the magnetoelectric effect is not taken into account there and one of the two forms of the Dzyaloshinsky-Moriya interaction used below is considered.

Electromagnons are a special aspect of the magnetization dynamics in multiferroics \cite{Pimenov NP 06}, \cite{ShuvaevPimenov EPJB 11}.
An example of observing electromagnons in the cycloidal magnetic phase is $TbMnO_{3}$
\cite{Aupiais npj QM 18}.
However, this type of perturbations is not considered upon in the results presented below.

\section{Spin-current model in systems of collinear spins}

The spin-current model is proposed in Ref. \cite{Katsura PRL 05} for the case when polarization occurs in systems of non-collinear spins.
Its generalization to the regime when polarization is formed due to the parallel component of spins is performed in Refs. \cite{AndreevTrukh JETP 24} and \cite{AndreevTrukh PS 24}.
Let us consider these results for the purpose of further development of this approach.

\subsection{Magnetoelectric effect for ferromagnets}

It is known that in systems of parallel spins or in the presence of their parallel component, the electric dipole moment arises in multiferroics
(see, for example, review \cite{Tokura RPP 14}, Fig. 2):
\begin{equation}\label{MFMemf edm operator simm}
\hat{\textbf{d}}_{ij}= \mbox{\boldmath $\Pi$}_{ij} (\hat{\textbf{s}}_{i}\cdot\hat{\textbf{s}}_{j}), \end{equation}
where the vector constant $\mbox{\boldmath $\Pi$}_{ij}$ is introduced.
As in Ref. \cite{Tokura RPP 14}, we assume that the constant $\mbox{\boldmath $\Pi$}_{ij}$ is a function of the distance between the spins.

Note that in the context of another mechanism for the formation of electric polarization associated with perpendicular components of spins,
there is the spin-current model \cite{Katsura PRL 07}
(it is also discussed in the review \cite{Tokura RPP 14}, see also Fig. 2 and equation 14).
Formally, this model comes down to the fact that the polarization of the medium $P^{\mu}$ is proportional to the spin current tensor $P^{\mu}
\sim\varepsilon^{\mu\alpha\beta}J^{\alpha\beta}$,
where $J^{\alpha\beta}$ is the spin current tensor, the first index refers to the spin, the second index refers to the momentum, and
$\varepsilon^{\mu\alpha\beta}$ is the absolutely antisymmetric unit tensor of the third rank (the Levi-Civita symbol).

In Refs. \cite{AndreevTrukh JETP 24}, \cite{AndreevTrukh PS 24} authors analyze the nature of the spin-current model.
The authors show that the balance of forces of the electric dipole-dipole interaction and the spin-orbit interaction
(the force acting on the moving magnetic ion (its magnetic moment) in the electric field of surrounding ions), existing both in the equilibrium state and in the presence of perturbations, leads to electric polarization of the medium in the form of
\begin{equation}\label{MFMemf spin current model} P^{\mu}
=\frac{\gamma}{c}\varepsilon^{\mu\alpha\beta}J^{\alpha\beta},
\end{equation}
where $\gamma$ is the gyromagnetic ratio, $c$ is the speed of light in the vacuum.
In the works \cite{AndreevTrukh JETP 24}, \cite{AndreevTrukh PS 24}, the form of the proportionality coefficient in the spin-current model is established.

The spin-current model (\ref{MFMemf spin current model}) does not imply a specific type of spin current.
Thus, spin currents of different nature can lead to the formation of polarization of the medium.
Using the effective spin current due to the Heisenberg exchange interaction leads to polarization due to non-collinear spins
\cite{Tokura RPP 14} (see eqs. 15-18), \cite{Katsura PRL 07}.
Spin current of magnons not accompanied by ion movement.
The authors of Ref. \cite{Katsura PRL 07} consider the spin current associated with individual ions
and obtain the electric dipole moment associated with the pair of neighboring magnetic ions.
The authors of Refs. \cite{AndreevTrukh JETP 24}, \cite{AndreevTrukh PS 24} consider the macroscopic model,
obtain the polarization of the medium, which is known from Refs. \cite{Sparavigna PRB 94}, \cite{Mostovoy PRL 06},
and establish the electric dipole moment, leading to the obtained macroscopic expression.

However, consideration of the effective spin current due to the Dzyaloshinskii-Moriya interaction leads to polarization due to collinear spins \cite{AndreevTrukh JETP 24}, \cite{AndreevTrukh PS 24}.
Note that the Dzyaloshinskii-Moriya interaction
$\hat{H}=(-1/2)\sum_{i,j,j\neq i}\textbf{D}_{ij}\cdot[\hat{\textbf{S}}_{i}\times \hat{\textbf{S}}_{j}]$
can be due to two mechanisms and therefore the Dzyaloshinskii constant consists of two terms
\cite{Fishman PRB 19} (see equation 14),
\cite{Khomskii JETP 21} (see text after equation 2):
\begin{equation}\label{MFMemf D Dz const str}
\textbf{D}_{ij}=\gamma(r_{ij})\textbf{r}_{ij}
+\beta(r_{ij})[\textbf{r}_{ij}\times\mbox{\boldmath $\delta$}], \end{equation}
where $\gamma(r_{ij})$ and $\beta(r_{ij})$ are isotropic functions of the distance between ions,
$\mbox{\boldmath $\delta$}$ is the displacement vector of the non-magnetic ligand ion (most often this is an oxygen ion) relative to the line of location of two neighboring magnetic ions.
In Refs. \cite{AndreevTrukh JETP 24} and \cite{AndreevTrukh PS 24} the second part of the Dzyaloshinskii constant associated with the ligand ion is considered.
The first part leads to a symmetric spin current and does not induce polarization of the medium.
As the result, the polarization of the medium of the form
\cite{AndreevTrukh JETP 24} (see equation 13, the moment of force and the corresponding spin current are presented by equations 9 and 10),
\cite{AndreevTrukh PS 24} (see equation 17, the moment of force and the corresponding spin current are presented by equations 36, 44 and 39, 45):
\begin{equation}\label{MFMemf P appr Symm}
\textbf{P}(\textbf{r},t)= \mbox{\boldmath $\delta$}
[ c_{0}(\textbf{S}\cdot\textbf{S})+c_{2}(\textbf{S}\cdot\triangle\textbf{S})]
, \end{equation}
arises, which, in particular, is proportional to the displacement vector of the nonmagnetic ion of the ligand included in the Dzyaloshinsky constant
$\mbox{\boldmath $\delta$}$.
Here, two constants
$c_{0}$ and $c_{2}$ arise. Let us present the relationship of the constants
$c_{0}$ and $c_{2}$ with the vector coefficient in the dipole moment
(\ref{MFMemf edm operator simm}) and the coefficient
$\beta$ included in the Dzyaloshinsky constant
(\ref{MFMemf D Dz const str}). The constants
$c_{0}$ and $c_{2}$ arise as integrals of the function
$\mbox{\boldmath $\Pi$}_{ij}$:
$c_{0}\mbox{\boldmath $\delta$}=\int \mbox{\boldmath $\Pi$}_{ij}(r_{ij})d^{3}r_{ij}$
and
$c_{2}\mbox{\boldmath $\delta$}=(1/6)\int r_{ij}^{2}\mbox{\boldmath $\Pi$}_{ij}(r_{ij})d^{3}r_{ij}$
and can be considered as moments of the function
$\beta(r)$ determining the type of the constant
$\textbf{D}_{ij}=\beta(r_{ij})[\textbf{r}_{ij}\times\mbox{\boldmath $\delta$}]$ (see equation 17 in
\cite{AndreevTrukh PS 24}). Let us represent the relationship of the function
$\mbox{\boldmath $\Pi$}_{ij}(r_{ij})$ from
\cite{Tokura RPP 14} (see Fig. 2) and the function
$\beta(r)$ included in the structure of the constant
$\frac{\partial \mbox{\boldmath $\Pi$}}{\partial r}
=\frac{\gamma}{c}r\beta(r)\mbox{\boldmath $\delta$}$ or a simplified form that is valid for the zeroth order
$\mbox{\boldmath $\Pi$}
=-\frac{\gamma}{3c}r^2\beta(r)\mbox{\boldmath $\delta$}$.
Note that the first term, proportional to
$c_{0}$, which, in the form of free energy, can be found in the review
\cite{Dong AinP 15} (see equation 9 at
$\gamma=1$), does not contribute to the moment of force.
Only the second term contributes to the magnetoelectric effect in the Landau--Lifshitz--Gilbert equation.

\subsection{Magnetoelectric effect for an antiferromagnet}

When considering an antiferromagnet of two types of atoms/ions $A$ and $B$, we choose their alternating arrangement $ABAB$ and, as for ferromagnets, use the formula (\ref{MFMemf edm operator simm}) for the electric dipole moment associated with a pair of neighboring spins, which, naturally, belong to different types of particles.

The momentum balance equation (analogous to the Navier-Stokes equation),
from which the relationship between the polarization of the medium and the spin current (\ref{MFMemf spin current model}) arises,
is derived for each type of atom separately.
As the result, expressions for partial polarizations $\textbf{P}_{A}$ and $\textbf{P}_{B}$ arise (see eqs 22 and 23 in Ref. \cite{AndreevTrukh PS 24}).
The total polarization determines the dynamics of the system through the energy density
$\mathcal{E}=-((\textbf{P}_{A}+\textbf{P}_{B})\cdot \textbf{E})=-(\textbf{P}\cdot \textbf{E})$
additively with respect to the partial polarizations.
We present it in the following work \cite{AndreevTrukh PS 24} (see eq. 24)
\begin{equation}\label{MFMafmUEI P def expanded AFR}
\textbf{P}=\mbox{\boldmath $\delta$}[2 c_{0,AB}
(\textbf{S}_{A}\cdot\textbf{S}_{B})
+c_{2,AB}
(S^{\nu}_{A} \triangle S^{\nu}_{B} +S^{\nu}_{B} \triangle S^{\nu}_{A})].
\end{equation}

\section{Model: Macroscopic Landau--Lifshitz--Gilbert equation}

\subsection{Ferromagnet model and polarization contribution to the moment of force}

The expression obtained above for the polarization (\ref{MFMemf P appr Symm}) can be used to construct the corresponding energy density $\mathcal{E}=-(\textbf{P}\cdot \textbf{E})$ and further obtain the contribution of the magnetoelectric effect to the equation of the evolution of the spin density $\partial_{t}\textbf{S}\mid_{\textbf{E}}=\gamma[\textbf{S}\times\textbf{H}_{eff}]$ by calculating the variational derivative \begin{equation}\label{MFMemf H eff caused by el f}\textbf{H}_{eff}=-\frac{1}{\gamma}\frac{\delta \mathcal{E}}{\delta \textbf{S}}
=-\frac{1}{\gamma}\biggl(\frac{\partial \mathcal{E}}{\partial \textbf{S}}
-\partial_{\beta}\frac{\partial \mathcal{E}}{\partial (\partial_{\beta}\textbf{S})}
+\triangle\frac{\partial \mathcal{E}}{\partial (\triangle\textbf{S})}\biggr) \end{equation}
up to the derivative with respect to the Laplacian of the spin density $\triangle\textbf{S}$,
since the energy density contains $\triangle\textbf{S}$ through the polarization of the medium (\ref{MFMemf P appr Symm}).

For a convenient analysis of the antiferromagnetic case, we present the Landau--Lifshitz--Gilbert equation for ferromagnets
$$\partial_{t}\textbf{S}=
A[\textbf{S}\times\triangle\textbf{S}]
+\kappa [\textbf{S}\times S_{z}\textbf{e}_{z}]
+\frac{1}{3}g_{(\gamma)}[\textbf{S}\times curl\textbf{S}]$$
$$
+\frac{1}{3}g_{(\beta)}\biggl((\textbf{S}\cdot[\mbox{\boldmath $\delta$}\times\nabla])\textbf{S}
-\frac{1}{2}[\mbox{\boldmath $\delta$}\times\nabla]S^{2}\biggr)
+2 c_{2}\varepsilon^{\alpha\beta\gamma}S^{\beta}\times$$
\begin{equation}\label{MFMemf s evolution MAIN TEXT}
\times
[(\mbox{\boldmath $\delta$}\cdot \textbf{E})\triangle S^{\gamma}
+(\mbox{\boldmath $\delta$}\cdot(\partial^{\delta} \textbf{E}))\cdot\partial^{\delta} S^{\gamma}]
+a[\textbf{S}\times\partial_{t}\textbf{S}] \end{equation}
which accordingly contains the isotropic exchange term, the contribution of the anisotropy energy due to the non-diagonality of the Heisenberg Hamiltonian (i.e., using the $X,Y,Z$ model at the microscopic level), the Dzyaloshinskii-Moriya interaction (consisting of two parts corresponding to two parts of the Dzyaloshinskii constant), the contribution of the magnetoelectric effect, and accounting for attenuation in the form proposed by Gilbert.

The first term on the right-hand side of the equation
describes the exchange interaction defined at the microscopic level by the Heisenberg Hamiltonian.
Its coefficient $A$ is the integral
of the exchange integral $J_{0}(r)$ considered as the function of the relative distance of the interacting magnetic ions
$A=\int r^2 J_{0}(r)d^{3}r/6$.
The second term describes the contribution of the anisotropy energy of the uniaxial crystal, with the anisotropy axis directed parallel to the z-axis.
Its coefficient kappa is related to the additional contribution to the Heisenberg Hamiltonian in the X, Y, Z model in the uniaxial regime
$H=J_{0}\textbf{S}_{1}\cdot\textbf{S}_{2}$$+\tilde{\kappa} S_{1}^{z}S_{2}^{z}$,
where
$\tilde{\kappa}(r)\equiv J_{zz}(r)-J_{0}(r)$,
and $\kappa=\int \tilde{\kappa}(r)d^{3}r$.
The terms describing the Dzyaloshinsky-Moriya interaction
contain coefficients
$g_{(\gamma)}=\int r^2 \gamma(r)d^{3}r$
and
$g_{(\beta)}=\int r^2 \beta(r)d^{3}r$
containing functions from the Dzyaloshinsky constant (\ref{MFMemf D Dz const str}).
The last term describes the Hilbert damping coefficient $a<0$.
The term proportional to $c_{2}$ describes the magnetoelectric effect discussed above
and the coefficient $c_{2}$ is described after the equation (\ref{MFMemf P appr Symm}).

For comparison, we present the polarization of the medium caused by the system of non-collinear spins (their perpendicular components)
\cite{Sparavigna PRB 94} and \cite{Mostovoy PRL 06}
\begin{equation}\label{MFMemf P def expanded} \textbf{P}(\textbf{r},t)=
\sigma
[\textbf{S}(\nabla\cdot \textbf{S})-(\textbf{S}\cdot\nabla)\textbf{S}], \end{equation}
and the corresponding moment of forces giving the magnetoelectric effect for systems of non-collinear spins
\cite{Risinggard SR 16} and \cite{Andreev 2025 05}
$$\textbf{T}=-\sigma
\biggl[ [\textbf{E}\times \nabla] S^{2}
-2(\textbf{S}\cdot[\textbf{E}\times\nabla]) \textbf{S}$$
\begin{equation}\label{MFMemf Torque magEl nonCol}-S^{2}(\nabla\times\textbf{E})
+\textbf{S}(\textbf{S}\cdot [\nabla\times\textbf{E}])
\biggr].
\end{equation}
The coefficient $\sigma$ can be represented via the exchange integral \cite{AndreevTrukh PS 24}
or via the energy difference and hybridization between the p - orbital and d - orbital (see eq. 20 \cite{Tokura RPP 14}).

A comparison of the equations (\ref{MFMemf s evolution MAIN TEXT}) and (\ref{MFMemf Torque magEl nonCol}) shows the features of the manifestation of the magnetoelectric effect in systems with collinear and non-collinear spins.

\subsection{Antiferromagnet model}

The macroscopic Landau--Lifshitz--Gilbert equations for the two-component antiferromagnet are written for the vectors
$\textbf{L}=\textbf{S}_{A}-\textbf{S}_{B}$ and
$\textbf{M}=\textbf{S}_{A}+\textbf{S}_{B}$.
For $\textbf{L}$, the equation has the following form:
$$\partial_{t}L^{\alpha}=-g_{0,u}\varepsilon^{\alpha\beta\gamma}L^{\beta}M^{\gamma}
+\kappa\varepsilon^{\alpha\beta z}M^{\beta}L^{z}
+A\varepsilon^{\alpha\beta\gamma}M^{\beta}\triangle L^{\gamma}$$
$$+\frac{1}{3}\varepsilon^{\alpha\beta\gamma}\varepsilon^{\gamma\mu\nu}\biggl[
M^{\beta}\hat{g}L^{\nu}\biggr]
+2 c_{0}\varepsilon^{\alpha\beta\gamma}M^{\beta}
(\mbox{\boldmath $\delta$}\cdot \textbf{E}) L^{\gamma}
+2 c_{2}\varepsilon^{\alpha\beta\gamma}M^{\beta}$$
\begin{equation}\label{MFMemf L evolution MAIN TEXT}
\times
[(\mbox{\boldmath $\delta$}\cdot \textbf{E})\triangle L^{\gamma}
+(\mbox{\boldmath $\delta$}\cdot(\partial^{\delta} \textbf{E}))\cdot\partial^{\delta} L^{\gamma}]
+a\varepsilon^{\alpha\beta\gamma}M^{\beta}\partial_{t}L^{\gamma}\end{equation}
where
$g_{0,u,AB}=-g_{0,u,AA}=-g_{0,u}$ and
$\hat{g}=g_{(\gamma)}\partial^{\mu}
+g_{(\beta)}\varepsilon^{\mu\delta\lambda}\delta^{\lambda}\partial^{\delta}$, and the approximation
$c_{2,AB}=-c_{2,AA}=-c_{2,BB}\equiv -c_{2}$,
$A\equiv g_{2,u}/6$ is also used.
We present the second equation of the system
$$\partial_{t}M^{\alpha}=\kappa\varepsilon^{\alpha\beta z}L^{\beta}L^{z}
+A\varepsilon^{\alpha\beta\gamma}L^{\beta}\triangle L^{\gamma}$$
$$
+\frac{1}{3}\varepsilon^{\alpha\beta\gamma}\varepsilon^{\gamma\mu\nu}\biggl[
L^{\beta}\hat{g}L^{\nu}
\biggr] +2 c_{2}\varepsilon^{\alpha\beta\gamma}L^{\beta}\times$$
\begin{equation}\label{MFMemf M evolution MAIN TEXT}
\times[(\mbox{\boldmath $\delta$}\cdot \textbf{E})\triangle L^{\gamma}
+(\mbox{\boldmath $\delta$}\cdot(\partial^{\delta} \textbf{E}))\cdot\partial^{\delta} L^{\gamma}]
+a\varepsilon^{\alpha\beta\gamma}L^{\beta}\partial_{t}L^{\gamma}.\end{equation}
When deriving the contribution of the magnetoelectric effect to the equations
(\ref{MFMemf L evolution MAIN TEXT}) and (\ref{MFMemf M evolution MAIN TEXT})
we used the expression for polarization
(\ref{MFMafmUEI P def expanded AFR})
and the procedure of varying the energy density (\ref{MFMemf H eff caused by el f}).
This derivation is discussed in more detail in the Appendix.

\section{Spin waves in cycloidal antiferromagnets}

\subsection{The case of collinear spins}

For the simple interpretation of the dispersion dependence of spin waves propagating in structures with the periodic order, we present the contribution of the magnetoelectric effect and compare the contributions of two types of DMI in the case when the equilibrium spins are collinear.

\subsubsection{Easy axis}

The anisotropy axis is parallel to the $z$ axis.
Let us write the equilibrium state in the form
 $\textbf{L}_{0}=L_{0}\textbf{e}_{z}$, $\textbf{M}_{0}=0$, $L_{0}=const$.
For small-amplitude perturbations
$\textbf{L}=\textbf{L}_{0}+\delta \textbf{L}$, $\textbf{M}=0+\delta \textbf{M}$
propagating along the x axis
$\delta \textbf{L}=\textbf{L}_{a}e^{-\imath\omega t+\imath k x}$,
$\delta \textbf{M}=\textbf{M}_{a}e^{-\imath\omega t+\imath k x}$,
we obtain that the coupled oscillations $\delta L_{x}\neq 0$ and $\delta L_{y}\neq 0$,
in the absence of DMI and MEE, give one branch of wave perturbations
$\omega^{2}=(\kappa+g_{0u})L_{0}^{2}Ak^2$ ($\kappa>0$, $g_{0u}>0$, $A>0$).
Whereas DMI removes the degeneracy of
$\omega^{2}=(\kappa+g_{0u})L_{0}^{2}[Ak^2\pm D k_{x}]$
and leads to two branches of the spectrum,
for each of which there is the linear contribution of the projection of the wave vector $k_{x}$.
The variable sign of the term $\pm D k_{x}$ can be compensated by the variable sign of the projection of the wave vector $k_{x}$.
However, for a fixed direction of propagation, a difference in the phase and group velocities of the two waves appears.
The perturbations $\delta L_{z}=0$ go to zero.

The simplest form of the spectrum of antiferromagnets, obtained for the easy-axis type of material $\omega^{2}=(\kappa+g_{0u})L_{0}^{2}Ak^2$,
can be compared with ferromagnets of the easy-plane type, since has the gapless spectrum
(when considering the minimal model including anisotropy energy and exchange interaction):
$\omega^{2}_{F}= AS_{0}^{2}k^2[\mid\kappa\mid+Ak^2]$ when replacing $\kappa+g_{0u}\rightarrow \mid\kappa\mid$.

Taking into account the magnetoelectric effect, we consider small-amplitude perturbations
$\textbf{L}=\textbf{L}_{0}+\delta \textbf{L}$, $\textbf{M}=0+\delta \textbf{M}$,
propagating in an arbitrary direction $\textbf{k}$:
$\delta \textbf{L}=\textbf{L}_{a}e^{-\imath\omega t+\imath \textbf{k}\cdot\textbf{r}}$,
$\delta \textbf{M}=\textbf{M}_{a}e^{-\imath\omega t+\imath \textbf{k}\cdot\textbf{r}}$.
We obtain the dispersion dependence
$$\omega^{2}=(\kappa+g_{0u}+2c_{0}(\mbox{\boldmath $\delta$}\cdot \textbf{E}_{0}))L_{0}^{2}\biggl[\kappa+(A+2c_{2}(\mbox{\boldmath $\delta$}\cdot \textbf{E}_{0}))k^2$$
\begin{equation}\label{MFMemf}
\pm \frac{1}{3}(g_{(\gamma)}k_{z}+g_{(\beta)}(\delta_{x}k_{y}-\delta_{y}k_{x}))+\imath\omega a\biggr], \end{equation}
$a<0$.
Here we can see the role of the magnetoelectric effect as the mechanism for controlling the exchange integral by changing the constant electric field.
Perturbations of the electric field are not considered here.

\emph{Easy plane regime}

The anisotropy axis is parallel to the $z$ axis.
Let us write the equilibrium state in the form $\textbf{L}_{0}=L_{0}\textbf{e}_{x}$, $\textbf{M}_{0}=0$, $L_{0}=const$.
For small-amplitude perturbations $\textbf{L}=\textbf{L}_{0}+\delta \textbf{L}$, $\textbf{M}=0+\delta \textbf{M}$,
without taking into account the DMI and the MEE,
we obtain $\delta L_{x}=0$, perturbations $\delta L_{y}\neq 0$ give the dispersion dependence of the form $\omega^{2}=g_{0u}AL_{0}^{2}k^2$,
perturbations $\delta L_{z}\neq 0$ give the dispersion dependence of the form
$\omega^{2}=g_{0u}L_{0}^{2}[\mid\kappa\mid+Ak^2]$ ($\kappa<0$, $g_{0u}>0$, $A>0$).
Note also that the easy-plane ferromagnets have the gapless spectrum
(when considering the minimal model including the anisotropy energy and the exchange interaction):
$\omega^{2}_{F}= AS_{0}^{2}k^2[\mid\kappa\mid+Ak^2]$ similar to perturbations $\delta L_{y}\neq 0$ when replacing $g_{0u}\rightarrow \mid\kappa\mid$.
However, perturbations $\delta L_{z}\neq 0$ are similar to the easy-axis ferromagnets,
where there is also the gap in the spectrum, although the shape of the spectrum differs $\omega_{F}=S_{0}[\kappa+Ak^2]$ ($\kappa>0$).


Taking into account the magnetoelectric effect and DMI,
we consider small-amplitude perturbations
$\textbf{L}=\textbf{L}_{0}+\delta \textbf{L}$, $\textbf{M}=0+\delta \textbf{M}$,
propagating in an arbitrary direction $\textbf{k}$
we obtain that DMI leads to hybridization of the spectra described above
\begin{equation}\label{MFMemf}
\left|
  \begin{array}{cc}
    \omega^{2} -\omega^{2}_{1}(\textbf{k}) & -\frac{\imath}{3}g_{D}g_{0u}L_{0}^{2} \\
    \frac{\imath}{3}g_{D}g_{0u}L_{0}^{2} & \omega^{2}-\omega^{2}_{2}(\textbf{k}) \\
  \end{array}
\right|=0,
\end{equation}
where
$g_{D}=g_{(\gamma)}k_{x}+g_{(\beta)}(\delta_{y}k_{z}-\delta_{z}k_{y})$,
$\omega^{2}_{1}(\textbf{k})=(g_{0u}+2c_{0}(\mbox{\boldmath $\delta$}\cdot \textbf{E}_{0}))
L_{0}^{2}(A+2c_{2}(\mbox{\boldmath $\delta$}\cdot \textbf{E}_{0}))k^2$,
$\omega^{2}_{2}(\textbf{k})=(g_{0u}+2c_{0}(\mbox{\boldmath $\delta$}\cdot \textbf{E}_{0}))
L_{0}^{2}[\mid\kappa\mid+(A+2c_{2}(\mbox{\boldmath $\delta$}\cdot\textbf{E}_{0}))k^2]$.

Note that in this case the DMI manifests itself in the even manner,
i.e. its contribution does not depend on whether the $x$ or $y$ or $z$ axis is in the positive or negative direction (but the sign of the projection will give a contribution when propagating at an angle to one of the axes):
\begin{equation}\label{MFMemf}
\omega^{4}-\omega^{2}(\omega^{2}_{1}(\textbf{k})+\omega^{2}_{2}(\textbf{k}))+\omega^{2}_{1}(\textbf{k})\omega^{2}_{2}(\textbf{k})
-\frac{1}{9}g_{D}^{2}g_{0u}^{2}L_{0}^{4}=0.
\end{equation}

\subsection{Cycloidal equilibrium order}

Let us consider the cycloidal equilibrium order
\begin{equation}\label{MFMemf cycloid for L}
\textbf{L}_{0}=L_{b}\cos(qx)\textbf{e}_{x}+L_{c}\sin(qx)\textbf{e}_{y},
\end{equation}
$\textbf{M}_{0}=0$
the conditions of its possibility,
within the framework of the given model, and then consider its perturbations.

Equilibrium of the presented type requires consideration of substances that lack one of the DMI types
$g_{2(\gamma)}=0$.
The second type of DMI, containing the ligand displacement vector, allows us to obtain the cycloid of the given type at
$\mbox{\boldmath $\delta$}\parallel \textbf{e}_{y}$.
Moreover, the coefficients of the cycloid must be related as  $L_{c}=\pm L_{b}$.
As a consequence of this, we obtain expressions for the derivatives of the projections of the equilibrium spin density
$\partial_{x}L_{0x}=\mp qL_{0y}$,
$\partial_{x}L_{0y}=\pm qL_{0x}$,
and
$L_{0x}^{2}+L_{0y}^{2}=L_{b}^{2}=const$.

Let us present the linearized Landau--Lifshitz--Gilbert equations for the two-component antiferromagnets
$$-\imath\omega\delta L^{\alpha}=\varepsilon^{\alpha\beta\gamma}\delta M^{\beta}
\biggl[(g_{0u}-Aq^{2}-2(c_{0}+c_{2}q^{2})\delta\cdot E_{0y})L_{0}^{\gamma}(x)$$
\begin{equation}\label{MFMemf} -\frac{1}{3}\varepsilon^{\gamma z\nu}g_{(\beta)}\delta\cdot\partial_{x}L_{0}^{\nu}(x)\biggr],
\end{equation}
and
$$-\imath\omega\delta M^{\beta}=\varepsilon^{\beta\sigma\lambda}L_{0}^{\sigma}(x)
\biggl[\kappa\delta^{\lambda z}\delta L_{z} +Aq^{2}\delta L^{\lambda} +A\triangle\delta L^{\lambda}
$$
$$-\frac{1}{3}\varepsilon^{\lambda z \mu}g_{(\beta)}\delta\cdot\partial_{x}\delta L^{\mu}
+2c_{2}\biggl(\delta \cdot E_{0y} \triangle\delta L^{\lambda}
+\partial^{\delta}(\delta\cdot\delta E_{y})\cdot\partial^{\delta}L_{0}^{\gamma}$$
\begin{equation}\label{MFMemf}
+q^{2}(\delta \cdot E_{0y})\delta L^{\lambda}\biggr)-\imath\omega a \delta L^{\lambda}\biggr]
-\frac{1}{3}\varepsilon^{\beta\sigma\lambda}\varepsilon^{\lambda z \mu}g_{(\beta)}\delta\cdot\partial_{x}L_{0}^{\mu}(x).
\end{equation}

Perturbations $\delta L_{z}\neq 0$ give the dispersion dependence of the form (without the IEE and Gilbert damping)
$\omega^{2}=(g_{0u}-Aq^2\pm D q)L_{b}^{2}[\mid\kappa\mid+A(k^2-q^2) \pm D q]$ ($\kappa<0$, $g_{0u}>0$, $A>0$, $D=g_{(\beta)}\delta/3$).
In the limiting case
$q\rightarrow0$ and $L_{b}\rightarrow L_{0}$
we obtain the transition to the collinear regime.

Next, taking into account the contribution of the magnetoelectric effect to the dynamics of $\delta L_{z}$, we obtain that the static electric field contributes to the dispersion dependence, leading to an effective shift of the exchange interaction constant, as in the homogeneous case:
$$\omega^{2}\delta L_{z}=L_{b}^{2}\biggl(g_{0u}+2c_{0}\delta\cdot E_{0y}-(A+2c_{2}\delta\cdot E_{0y})q^2\pm \frac{1}{3}g_{(\beta)}q\delta\biggr)\times$$
$$\times\biggl[\biggl(-\kappa -(A+2c_{2}\delta\cdot E_{0y})q^2\pm \frac{1}{3}g_{(\beta)}q\delta$$
\begin{equation}\label{MFMemf}
+\imath\omega a\biggr)\delta L_{z}
-(A+2c_{2}\delta\cdot E_{0y})\triangle\delta L_{z}\biggr]
.\end{equation}
Since this is a linear differential equation with constant coefficients, we can use the Fourier transform with respect to the spatial coordinate (replace the Laplacian $\triangle$ with $-k^2$)
and reduce by $\delta L_{z}$ to obtain the dispersion dependence.
Note that this equation without variable coefficients is obtained from the linearized equations by algebraic transformations without additional simplifications.
The stability of the system requires that the anisotropy constant $\kappa<0$ be negative, in accordance with the homogeneous case.
However, the contribution of the non-collinear equilibrium order leads to a decrease in the minimum frequency arising at $k=0$.
It is evident that the dynamic part of the electric field does not affect $\delta L_{z}$
(it is shown below that $\delta L_{z}$ does not contribute to the polarization perturbations).
Thus, the presented mode does not contribute to the response of the medium
(for equilibrium in the form of a flat cycloid, for conical structures the situation is likely to change).

Let us proceed to consider the coupled dynamics $\delta L_{x}$ and $\delta L_{y}$.
To obtain an equation with constant coefficients and the subsequent derivation of the dispersion dependence, after transforming the coefficients of the corresponding equations, we introduce an auxiliary function
\begin{equation}\label{MFMemf}
\delta f= L_{0y}\delta L_{x}-L_{0x}\delta L_{y}
. \end{equation}
Identical transformations allow us to obtain the following equation for this function (and the perturbation of the electric field)
$$\omega^{2}\delta f= L_{b}^{2}\biggl(g_{0u}+2c_{0}\delta\cdot E_{0y}-(A+2c_{2}\delta\cdot E_{0y})q^2\pm \frac{1}{3}g_{(\beta)}q\delta\biggr)\times$$
\begin{equation}\label{MFMemf}
\times[-(A+2c_{2}\delta\cdot E_{0y})\triangle\delta f +\imath\omega a\delta f \pm2c_{2}q\delta \cdot L_{b}^{2}\partial_{x}\delta E_{y}]
. \end{equation}
The resulting equation with constant coefficients can be reduced to an algebraic equation using the Fourier transform.
Neglecting the contribution of the perturbation of the electric field, we obtain the dispersion equation for spin waves:
$$\omega^{2}=L_{b}^{2}\biggl(g_{0u}+2c_{0}\delta\cdot E_{0y}-(A+2c_{2}\delta\cdot E_{0y})q^2\pm \frac{1}{3}g_{(\beta)}q\delta\biggr)\times$$
\begin{equation}\label{MFMemf DE LxLy C}
\times((A+2c_{2}\delta\cdot E_{0y})k^2 +\imath\omega a)
. \end{equation}
This equation allows us to obtain the dispersion dependence of spin waves $\omega_{0f}(k)$.
Taking into account the perturbation of the electric field, we can obtain the dispersion dependence of coupled electromagnetic and spin waves, similar to the works of \cite{Bychkov SSP 12} and \cite{Bychkov JMMM 13}.
However, below we limit ourselves to obtaining the dynamic complex tensor of permittivity (in our case, it is reduced to a single element $\varepsilon_{yy}$).

In the equation (\ref{MFMemf DE LxLy C}) it is clear that the constant electric field modifies the exchange integral $(A+2c_{2}\delta\cdot E_{0y})$, and the spiral equilibrium structure affects the type of the coefficient $g_{0u}$, leading to its modification, both due to the exchange interaction and due to the DMI.

Note that if $g_{(\gamma)}$ is neglected and the direction of wave propagation is chosen along the $Ox$ axis, the two wave modes do not influence each other,
in accordance with the homogeneous case considered above.

\section{Dielectric constant of multiferroics}

Consideration of antiferromagnets with the $AABB$ configuration when obtaining the contribution of the magnetoelectric effect to the spin evolution equation
requires the use of polarization of the form $\textbf{P}_{\sum}=\textbf{P}_{AA}+\textbf{P}_{BB}+\textbf{P}_{AB}$.
Using the $c_{0,AB}=-c_{0,AA}=-c_{0,BB}=-c_{0}$
and $c_{2,AB}=-c_{2,AA}=-c_{2,BB}=-c_{2}$ approximations, the expression for the polarization $\textbf{P}_{\sum}$ can be simplified to $\textbf{P}_{\sum}=\mbox{\boldmath $\delta$}[c_{0}\textbf{L}^{2}+c_{2}\textbf{L}\cdot\triangle\textbf{L}]$.
We obtain an expression for the equilibrium polarization $\textbf{P}_{0,\sum}=\mbox{\boldmath $\delta$}\cdot L_{b}^{2}[c_{0}-q^2 c_{2}]$, and for its perturbation in the linear small-amplitude mode $\delta\textbf{P}_{\sum}=\mbox{\boldmath $\delta$}c_{2}(\textbf{L}_{0}\cdot\triangle\delta\textbf{L})$
$=\pm 2q\partial_{x} \delta f$.
It is expressed through the function $\delta f$ introduced above, which allows us to describe the perturbations $\delta L_{x}$ and $\delta L_{y}$.

We introduce the permittivity
$\delta P_{y,\sum}=(\varepsilon_{yy}-1)\delta E_{y}/4\pi$,
for which we obtain an explicit form:
\begin{equation}\label{MFMemf}
\varepsilon_{yy}=1-16\pi \tilde{g}_{0u}\frac{(c_{2}\delta\cdot qk_{x})^2 L_{b}^{4}}{\omega^{2}-\omega^{2}_{0f}},\end{equation}
where
$\tilde{g}_{0u}=\biggl(g_{0u}+2c_{0}\delta\cdot E_{0y}-(A+2c_{2}\delta\cdot E_{0y})q^2\pm \frac{1}{3}g_{(\beta)}q\delta\biggr)$.

Along with the fact that one of the two eigenwaves participates in the resonant response of the antiferromagnet,
we have found that the constant component of the external electric field allows us to vary the contribution of the exchange interaction.
Although two constants change simultaneously $g_{0u}+2c_{0}\delta\cdot E_{0y}$ and $A+2c_{2}\delta\cdot E_{0y}$.
Moreover, a large electric field can affect the values of the constants $g_{0u}$ and $A$ themselves, leading to superposition of the effects.
This behavior enables controlled manipulation of the system's properties through the application of an electric field in systems with the magnetoelectric effect under consideration.

\section{Conclusion}

A detailed discussion of the spin-current model of multiferroic polarization for systems of collinear spins (or when the main contribution to the polarization is due to parallel components of the spins) for ferromagnets and antiferromagnets is presented.
The corresponding expressions for the polarization of the medium are obtained.
The contribution of the magnetoelectric effect to the equations of the evolution of spin densities is considered.

Based on the presented model, the dispersion dependencies of spin waves are studied for the perturbations of the cycloidal order of the equilibrium spin density of antiferromagnets. The influence of two types of Dzyaloshinsky-Moriya interaction on the possibility of forming a cycloidal equilibrium order is discussed.
The dynamic permittivity as a function of frequency for such systems is calculated.

\section{DATA AVAILABILITY}
Data sharing is not applicable to this article as no new data were
created or analyzed in this study, which is a purely theoretical one.

\section{Acknowledgements}

The research is supported by the Russian Science Foundation under the
grant
No. 25-22-00064.

\appendix

\section{Contribution of the magnetoelectric effect to the spin evolution equations for antiferromagnets}

Let us consider in more detail the contribution of the magnetoelectric effect to the spin evolution equations in an antiferromagnet. In the case of the spin arrangement $ABAB$, the polarization associated with a pair of neighboring spins is determined by the coupling of ions of different types. Therefore, the polarization is determined by the equation (\ref{MFMafmUEI P def expanded AFR}).
Let us obtain the corresponding contribution to the spin evolution equation for particles of type $A$:
$\partial_{t}\textbf{S}_{A}\mid_{\textbf{E}}=\gamma[\textbf{S}_{A}\times\textbf{H}_{eff,A}]$
by calculating the variational derivative
$\textbf{H}_{eff,A}=-\frac{1}{\gamma}\frac{\delta \mathcal{E}}{\delta \textbf{S}_{A}}$
up to the derivative with respect to the Laplacian of the spin density $\triangle\textbf{S}$,
since the energy density contains $\triangle\textbf{S}_{A}$ through the polarization of the medium (\ref{MFMemf P appr Symm}).
As a result, we obtain
$$\partial_{t}S^{\alpha}_{A}\mid_{E}=2 c_{0,AB}\varepsilon^{\alpha\beta\gamma}S_{A}^{\beta}
[(\mbox{\boldmath $\delta$}\cdot \textbf{E}) S_{B}^{\gamma}$$
\begin{equation}\label{MFMemf s A evolution for AFM non simm ABAB}
+2 c_{2,AB}\varepsilon^{\alpha\beta\gamma}S_{A}^{\beta}
[(\mbox{\boldmath $\delta$}\cdot \textbf{E})\triangle S_{B}^{\gamma}
+(\mbox{\boldmath $\delta$}\cdot(\partial^{\delta} \textbf{E}))\cdot\partial^{\delta} S_{B}^{\gamma}]. \end{equation}

Let us move on to the variables $\textbf{L}=\textbf{S}_{A}-\textbf{S}_{B}$
and $\textbf{M}=\textbf{S}_{A}+\textbf{S}_{B}$.
For the vector $\textbf{L}$, we obtain
$$\partial_{t}L^{\alpha}\mid_{E}=c_{2,AB}\varepsilon^{\alpha\beta\gamma}L^{\beta}
[(\mbox{\boldmath $\delta$}\cdot \textbf{E})\triangle M^{\gamma}
+(\mbox{\boldmath $\delta$}\cdot(\partial^{\delta} \textbf{E}))\cdot\partial^{\delta} M^{\gamma}]$$
\begin{equation}\label{MFMemf L evolution for AFM non simm ABAB}
+c_{2,AB}\varepsilon^{\alpha\beta\gamma}M^{\beta}
[(\mbox{\boldmath $\delta$}\cdot \textbf{E})\triangle L^{\gamma}
+(\mbox{\boldmath $\delta$}\cdot(\partial^{\delta} \textbf{E}))\cdot\partial^{\delta} L^{\gamma}]. \end{equation}
Also, for the vector $\textbf{M}$, we obtain
$$\partial_{t}M^{\alpha}\mid_{E}=c_{2,AB}\varepsilon^{\alpha\beta\gamma}M^{\beta}
[(\mbox{\boldmath $\delta$}\cdot \textbf{E})\triangle M^{\gamma}
+(\mbox{\boldmath $\delta$}\cdot(\partial^{\delta} \textbf{E}))\cdot\partial^{\delta} M^{\gamma}]$$
\begin{equation}\label{MFMemf M evolution for AFM non simm ABAB}
+c_{2,AB}\varepsilon^{\alpha\beta\gamma}L^{\beta}
[(\mbox{\boldmath $\delta$}\cdot \textbf{E})\triangle L^{\gamma}
+(\mbox{\boldmath $\delta$}\cdot(\partial^{\delta} \textbf{E}))\cdot\partial^{\delta} L^{\gamma}]. \end{equation}

In the case of the spin arrangement $AABB$, the polarization associated with a pair of neighboring spins is determined by the coupling of both ions of the same type $AA$ and $BB$and ions of different types $AB$.
Therefore, for complete polarization, it is necessary to take into account the equation (\ref{MFMemf P def expanded}) for types  $AA$ and $BB$, along with the equation (\ref{MFMafmUEI P def expanded AFR}) considered above.
This will lead to an additional contribution
$$\partial_{t}\textbf{S}_{A}\mid_{E}=2 c_{0,AB}\varepsilon^{\alpha\beta\gamma}S_{A}^{\beta}
[(\mbox{\boldmath $\delta$}\cdot \textbf{E}) S_{B}^{\gamma}$$
$$+2c_{2,AA}\varepsilon^{\alpha\beta\gamma}S_{A}^{\beta}
[(\mbox{\boldmath $\delta$}\cdot \textbf{E})\triangle S_{A}^{\gamma}
+(\mbox{\boldmath $\delta$}\cdot(\partial^{\delta} \textbf{E}))\cdot\partial^{\delta} S_{A}^{\gamma}]
$$
\begin{equation}\label{MFMemf s A evolution for AFM non simm}
+2 c_{2,AB}\varepsilon^{\alpha\beta\gamma}S_{A}^{\beta}
[(\mbox{\boldmath $\delta$}\cdot \textbf{E})\triangle S_{B}^{\gamma}
+(\mbox{\boldmath $\delta$}\cdot(\partial^{\delta} \textbf{E}))\cdot\partial^{\delta} S_{B}^{\gamma}]. \end{equation}
However, this contribution (under the condition $c_{2,AB}=-c_{2,AA}=-c_{2,BB}\equiv -c_{2}$ ) leads to a simpler contribution of the magnetoelectric effect to the equations of evolution of vectors $\textbf{L}$ and $\textbf{M}$.
For the vector $\textbf{L}$ we obtain
$$\partial_{t}L^{\alpha}\mid_{E}=2 c_{0,AB}\varepsilon^{\alpha\beta\gamma}M^{\beta}
[(\mbox{\boldmath $\delta$}\cdot \textbf{E}) L^{\gamma}$$
\begin{equation}\label{MFMemf L evolution for AFM non simm AABB}
+2c_{2}\varepsilon^{\alpha\beta\gamma}M^{\beta}
[(\mbox{\boldmath $\delta$}\cdot \textbf{E})\triangle L^{\gamma}
+(\mbox{\boldmath $\delta$}\cdot(\partial^{\delta} \textbf{E}))\cdot\partial^{\delta} L^{\gamma}]. \end{equation}
Also, for the vector $\textbf{M}$ we obtain
\begin{equation}\label{MFMemf M evolution for AFM non simm AABB}
\partial_{t}M^{\alpha}\mid_{E}=2 c_{2}\varepsilon^{\alpha\beta\gamma}L^{\beta}
[(\mbox{\boldmath $\delta$}\cdot \textbf{E})\triangle L^{\gamma}
+(\mbox{\boldmath $\delta$}\cdot(\partial^{\delta} \textbf{E}))\cdot\partial^{\delta} L^{\gamma}]. \end{equation}

\end{document}